# How to Apply Assignment Methods that were Developed for Vehicular Traffic to Pedestrian Microsimulations

**Vidal Roca**
Regional Director, PTV Group, Germany
**Vicente Torres**
Managing Director, PTV América Latina, Mexico
**Tobias Kretz**
Chief Technical Product Manager PTV Viswalk, PTV Group, Germany
**Karsten Lehmann**
Software Developer, init AG, Germany
**Ingmar Hofsäß**
Senior Software Developer, PTV Group, Germany

**ABSTRACT**

Applying assignment methods to compute user-equilibrium route choice is very common in traffic planning. It is common sense that vehicular traffic arranges in a user-equilibrium based on generalized costs in which travel time is a major factor.

Surprisingly travel time has not received much attention for the route choice of pedestrians. In microscopic simulations of pedestrians the vastly dominating paradigm for the computation of the preferred walking direction is set into the direction of the (spatially) shortest path.

For situations where pedestrians have travel time as primary determinant for their walking behavior it would be desirable to also have an assignment method in pedestrian simulations.

To apply existing (road traffic) assignment methods with simulations of pedestrians one has to reduce the nondenumerably many possible pedestrian trajectories to a small subset of routes which represent the main, relevant, and significantly distinguished routing alternatives.

All except one of these routes will mark detours, i.e. not the shortest connection between origin and destination. The proposed assignment method is intended to work with common operational models of pedestrian dynamics. These – as mentioned before – usually send pedestrians into the direction of the spatially shortest path. Thus, all detouring routes have to be equipped with intermediate destinations, such that pedestrians can do a detour as a piecewise connection of segments on which they walk into the direction of the shortest path. One has then to take care that the transgression from one segment to the following one no artifacts are introduced into the pedestrian trajectory.

## 1. Introduction, Motivation, Task Formulation

As stated in the abstract we want to construct a method which computes relevant routing alternatives for a pedestrian microsimulation such that existing assignment methods – which were formulated to compute an route choice equilibrium in road traffic (Wardrop, 1952), (Beckmann, McGuire, & Winsten, 1956), (LeBlanc, Morlok, & Pierskalla, 1975), (Bar-

Gera, 2002), (Gentile & Nökel, 2009) i.e. on a graph structure – as well as existing models of operational pedestrian dynamics (Helbing & Molnar, 1995), (Burstedde, Klauck, Schadschneider, & Zittartz, 2001) (Pelechano, Allbeck, & Badler, 2007), (Schadschneider, Klüpfel, Kretz, Rogsch, & Seyfried, 2009), (Guy, et al., 2010), (Ondrej, Pettré, Olivier, & Donikian, 2010) – which send pedestrians into the direction of the spatially shortest path – can be used without having artifacts introduced into the pedestrian trajectories.

## 1.1. Artifacts in pedestrian trajectories

At first we have a look at the problem of artifacts when intermediate destinations are used to guide pedestrians, which in the simulation are heading for the (intermediate) destination into the direction of the shortest path, on a – under global perspective – detour to destination.

Imagine a walking geometry as shown in Figure 1. A pedestrian following the above mentioned principle of walking into the direction of the shortest path would pass the obstacle on the lower side (reader's perspective) as it is shown in the left figure. In this case it is easy to make pedestrians pass the obstacle on either side by introducing on each side of the obstacle an intermediate destination area. With the intermediate areas and the routes which include them, it is possible to route pedestrians locally into the direction of the shortest path, but still make a given fraction of pedestrians detour: first the pedestrians would head toward one of the two intermediate destinations and as soon as it is reached proceed to the final destination.

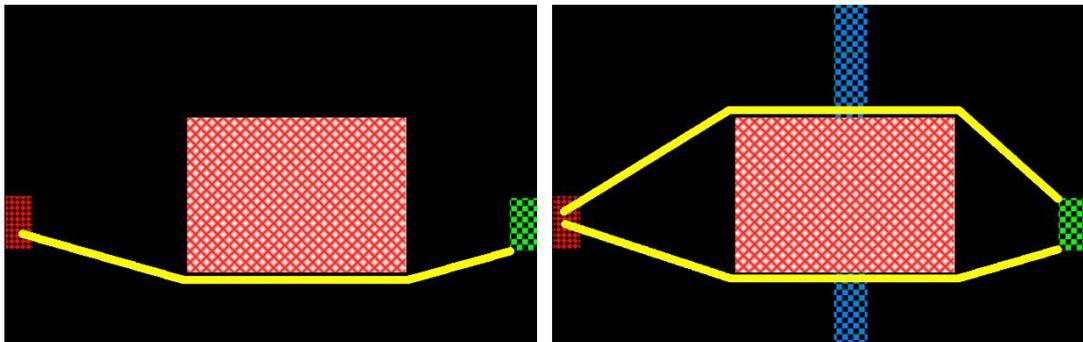

**Figure 1: Walking area (black), obstacle (diagonal red lines on white ground), origin area at the left sides (red and black diagonal checkerboard pattern) and destination at the right ends (green and black checkerboard pattern). Left figure: The yellow line shows the path a pedestrian (set into the simulation at some arbitrary coordinate on the origin area) would follow if a distance map is used to determine his basic direction. The route data simply would be some information (e.g. an area ID) identifying the destination area. Right figure: Compared to the left side figure the black-blue areas mark intermediate destination areas. There are two routes, one leading over each of the intermediate destination areas. (The intermediate destination is not necessary along the shorter (left) path, it has been added for illustration.)**

In Figure 2 the intermediate destination area on the left side was not necessary. Having a route directly leading from the origin to the destination area would have the same effect. However, it is important to note that the lower intermediate destination area does not change the path of the pedestrians on that route compared to the case without any intermediate destination area. This is not in general the case. In general it is difficult to shape the intermediate destinations such that the path on the principally shortest route is not distorted compared to the case without intermediate destination. Figure 3 shows such a case. It is concluded that the intermediate destination areas cannot be of trivial shape in general.

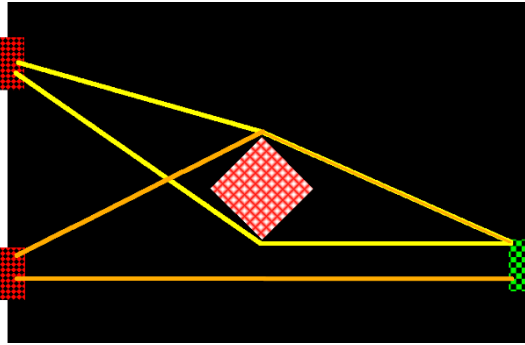

**Figure 2: Example with necessarily non-trivial geometry for the intermediate destination areas. Walking area (black), obstacle (diagonal red stripes on white ground), two origin areas to the left (red-black diagonal checkerboard), destination area to the right (green-black checkerboard), and shortest paths (yellow and orange).**

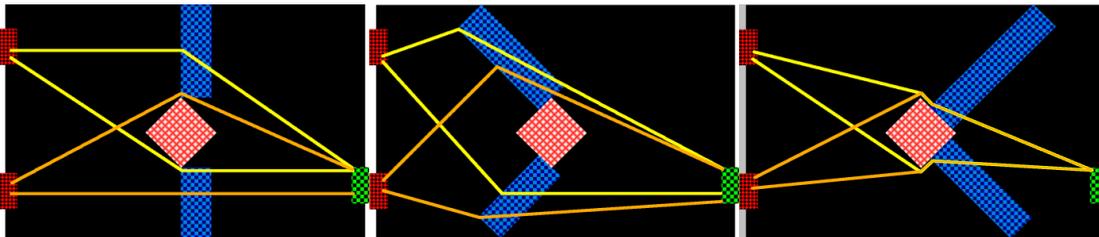

**Figure 3: For the example of Figure 2 simple rectangular areas are used as intermediate destination areas (light and dark blue checkerboard pattern). With none of the three variants the shortest paths as shown in Figure 2 are reproduced.**

The solution to the problem is to shape the upstream edge of the intermediate destinations along equi-distance lines to the next downstream (intermediate) destination. This means that the upstream edge is the set of points of which each has the same distance to the closest point of the downstream intermediate destination (considering obstacles when measuring distances). This is illustrated in Figure 4.

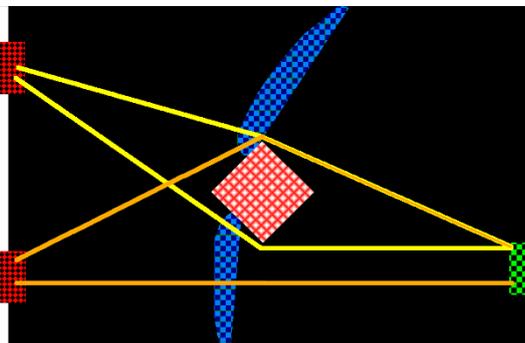

**Figure 4: The example of Figure 2 supplemented with two intermediate destination areas whose upstream edges are equi-distant lines to the destination area (green-black). It can be seen that all pedestrian trajectories enter the intermediate destination areas orthogonally. This is why pedestrian trajectories do not bend at the upstream edges, but remain straight (to be precise: differentiable).**

2. Computing Routes and Intermediate Destinations

Due to space limitations the method to compute the routes can only be sketched here. The full method has been published in (Kretz, Lehmann, & Hofsäss, 2014). We also refer to that paper for an extensive overview and discussion of previous relevant work.

As a first step a map of distances to destination is computed. In this map certain ranges of distance (for example each 2 meters) are combined to areas. The extent of this combination (2 or 5 or 10 meters) is a parameter of the method. The larger, the less alternatives will be found. A routing alternative is then found if there are two or more unconnected areas which have the same distance to destination. Figure 5 visualizes this idea. Once a routing alternative has been found one intermediate destination is created for each alternative. This process has to be repeated iteratively for each newly found intermediate destination until the entire walking area or at least all input areas are covered in an iteration step which does not bring up another routing alternative.

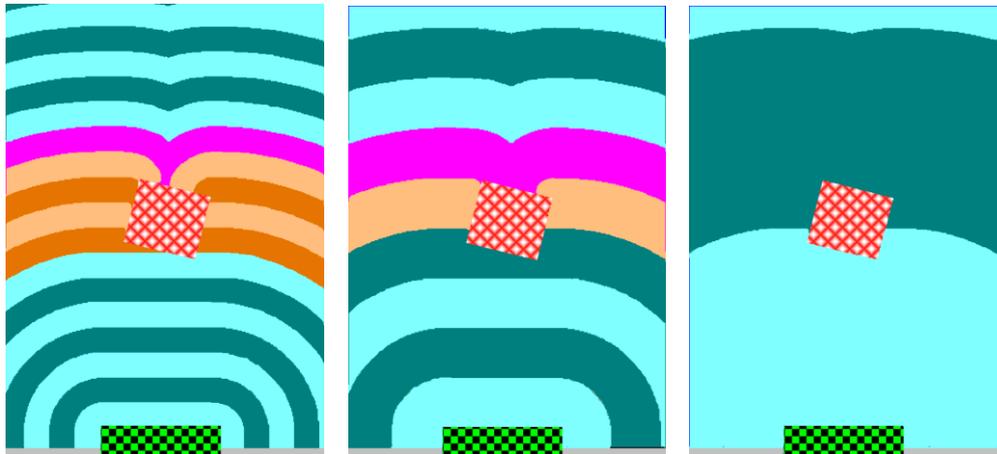

**Figure 5: The destination is shown as green and black checkerboard, an obstacle is colored with diagonal red lines on white ground. Light and dark cyan mark regions which are simply connected and which lie within a range of distances to destination and which have exactly one simply connected area as direct neighbor which is closer to destination. Light and dark orange mark regions where there is more than one (here: exactly two) unconnected (split by an obstacle) regions which are within a range of distances to destination. The magenta region is simply connected and has more than one (here: exactly two) directly neighboring areas which are closer to destination than the magenta region is itself.**

3. Example Application with an Assignment

The proposed method is now applied with an example scenario which is the same as case study 1 in (Hoogendoorn, Daamen, Duives, & van Wageningen-Kessels, 2014) or to be more precise in as much agreement as possible according to the information in the paper. Figure 6 shows the example and the intermediate destination areas and four routes which are calculated with the method proposed above. As base for the pedestrian simulation we have used PTV Viswalk (PTV Group, 2011) which is built on the combination of two variants (circular and elliptical II) of the Social Force Model (Johansson, Helbing, & Shukla, 2007). We have sticked with the Viswalk default parameters of the Social Force Model which especially implies that we have used the speed distributions as the International Maritime Organization defines them for men and women between 30 and 50 years (International Maritime Organization, 2007).

The four routes and the average travel times of these are used for an assignment. We do the assignment twice with two different initial conditions: a) equal load on each route and b) 97% load on route 4 (which is the shortest one) and 1% on each of the three other routes. The calculation of the new route choice ratios for the next iteration is done in a very simple

.

way. The probability to choose the route with the longest average travel time $t_{Max}$ is reduced by the same amount $\Delta p$ as the probability for the route with the smallest travel time $t_{Min}$ is increased:

$$\Delta p = 0.1 \left(\frac{t_{Max} - t_{Min}}{t_{Max} + t_{Min}}\right) \quad (1)$$

The process was terminated as soon as the average travel times on all routes with a load larger zero were in a range of 0.5 seconds.

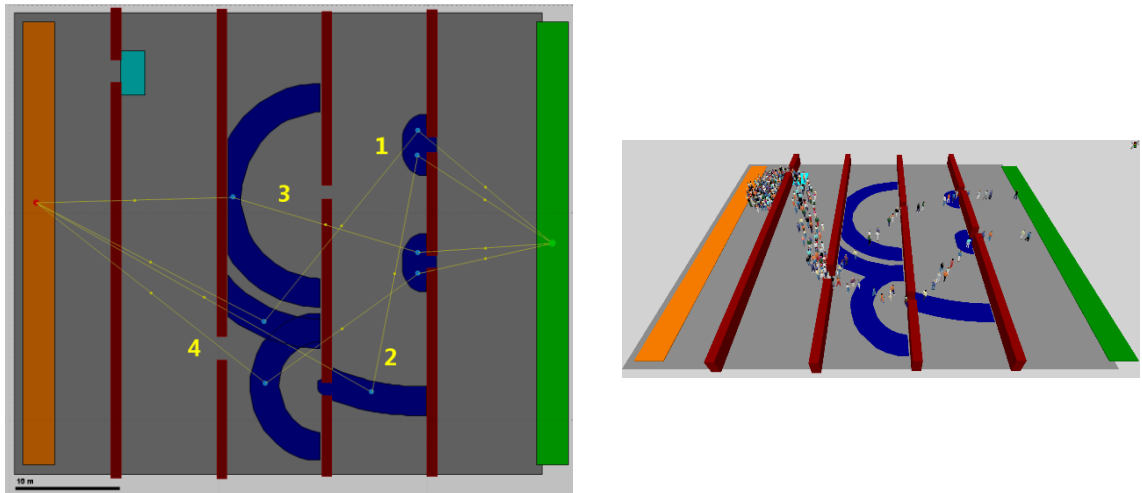

**Figure 6 (left side for information, right side for illustration of dimensions): Example scenario. 321 pedestrians (in this way the density on input area is 2.5 pedestrians per m²) at the beginning are set into the simulation and walk from the orange origin area on the left side to the green destination area on the right side; at later times no further pedestrians are added. Obstacles are shown dark red. The dark blue areas are intermediate destination areas which have been created by the method proposed in subsection 1.1 and section 2. The yellow lines mark the four routing alternatives which consist of a sequence of the origin area, two intermediate destination areas and the destination area. The cyan colored area marks the beginning of the travel time measurement. As soon as a pedestrian for the first time steps on the cyan area travel time measurement begins for him. It ends in the moment in which he arrives on the destination area. The average of these travel times is taken for the assignment process.**

It should be noted that for the assignment the average travel time on each route over the whole simulation is utilized. It can be argued that this is not fully realistic. A large fraction of early pedestrians might realistically walk along the shortest route (route 4), while later pedestrians might rather distribute according to available capacities. In a microsimulation one could in principle do the assignment separate for different time slices (here, however, all pedestrians begin their trip at the same time) or for different parts of the input area. It would, however, be problematic, if each population for which an assignment is done, is too small. If for example the starting area was divided into ten parts, each of these parts would contain on average 32 pedestrians. This implies that each pedestrian makes up for about 3% of the route choice ratio. The results would oscillate and convergence was questionable.

In the macroscopic approach of (Hoogendoorn, Daamen, Duives, & van Wageningen-Kessels, 2014) and (van Wageningen-Kessels, Daamen, & Hoogendoorn, 2014), but also in the microscopic one-shot assignment presented in (Kretz, 2009) and (Kretz, et al., 2011) and this is different. The implicit assumption there is that pedestrians can change their mind which route (or which door) they want to take. Particularly at the beginning and the end of the simulation this can yield more realistic behavior, however, with these approaches the route choice ratios do not easily result from the simulation respectively computation, but need to be restored or estimated from the resulting data.

Figure 7 and Figure 8 show the results for travel times and route choice ratios in the course of iterations. Note that the travel times are *average* travel times within the travel time measurement which begins not at simulation start but later and individually for each pedestrian and not the time for evacuation (time when the *last* pedestrian reaches the destination). In this case the result of the assignment is not identical with respect to the initial conditions. This is probably due to the relatively small number of pedestrians and the relatively large impact of the early and the late phase of the simulation as it is typical for evacuation simulations. The impact of the early and the late phase of a simulation can be excluded entirely if one continues to set pedestrians into the simulation. We have investigated such scenarios elsewhere (Kretz, Lehmann, & Friderich, 2013) (Kretz, Lehmann, & Hofsäss, 2014b) (Kretz, Lehmann, Hofsäß, & Leonhardt, 2014), but wanted to stick here with an evacuation setting to better compare to case study 1 of (Hoogendoorn, Daamen, Duives, & van Wageningen-Kessels, 2014).

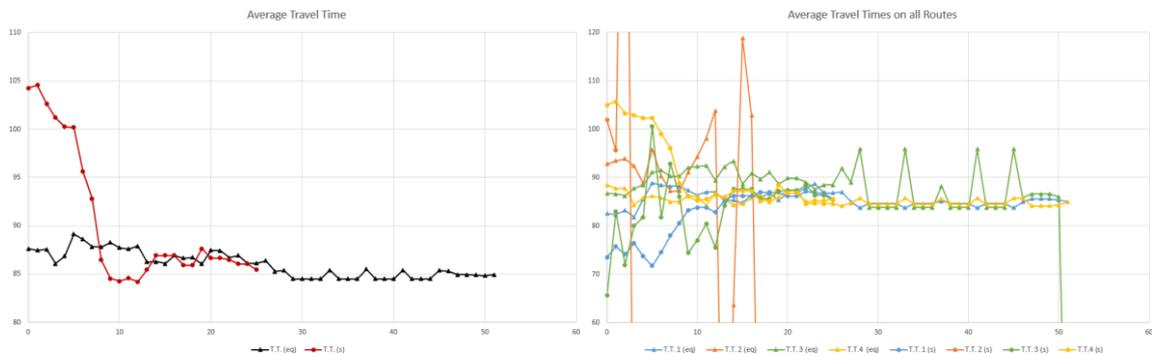

**Figure 7: Overall average travel time (left) and per route (right) in the course of iterations if in the first iteration all routes are equally utilized "(eq)" and if 97% of pedestrians are sent on the shortest route "(s)".**

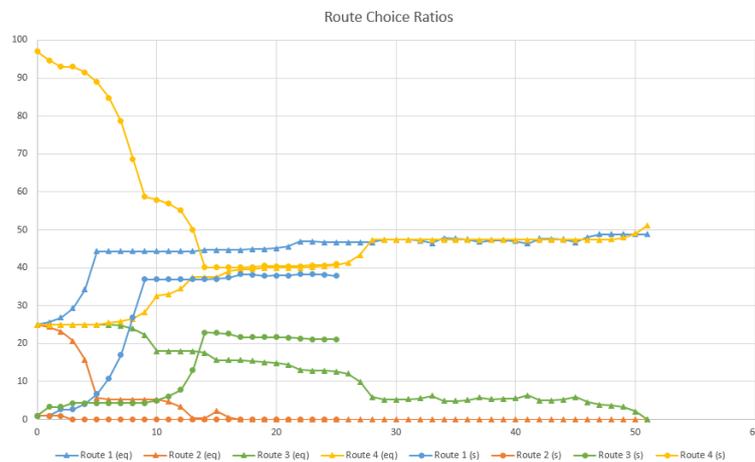

**Figure 8: Route choice ratios in the course of iterations.**

Figure 9 and Figure 10 show the density distribution at two times in the simulation as it results when pedestrians are not guided via intermediate destinations and therefore all walk along the shortest path and how this changes with the assignment. It can be seen in Figure

10 that the jams remain about the same when there is no alternative option for a door (the first two doors) but that it is significantly reduced when available alternative options are used.

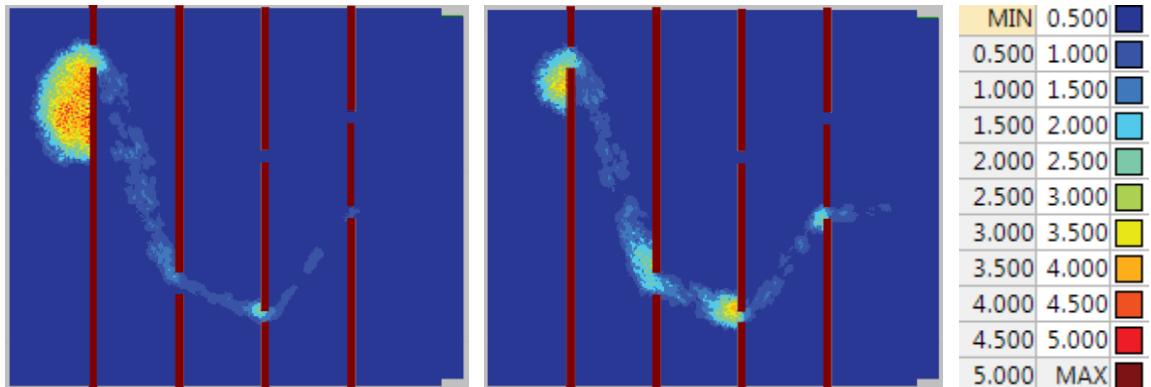

**Figure 9: Density distribution after 50 and after 125 seconds of simulation without assigning intermediate destinations to pedestrians and therefore effectively having all pedestrian walk along route 4 (shortest route). To compute the density a regular lattice was placed over the area with a spacing of 20 cm. Pedestrians in a vicinity of cells into each main direction contributed to the density (i.e. density was measured in approximate circles with a diameter of 2.2 m). Furthermore the density was averaged over 1 second (10 simulation time steps).**

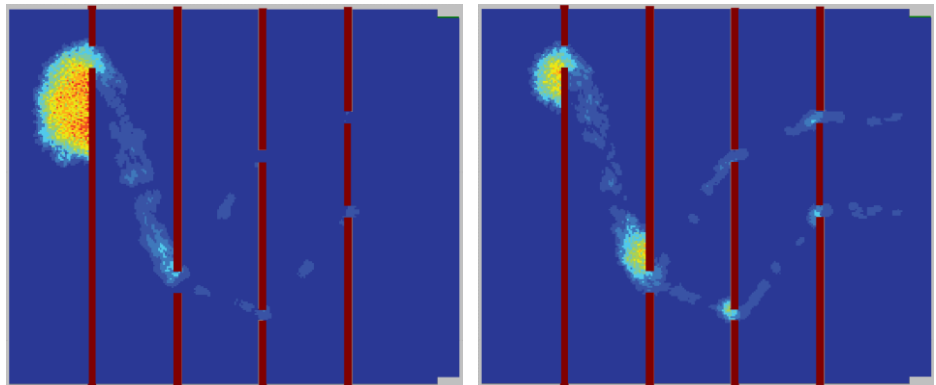

**Figure 10: Density distribution after the assignment. Now pedestrians also utilize other doors and routes.**

Table 1 shows a comparison of global travel times (note: global travel times last from the time of creation of a pedestrian to the time when he reaches the destination which is longer than the travel times which were relevant for the assignment) in different route configurations. "One-shot assignment" means that no intermediate destinations are assigned to the pedestrians, but pedestrians in each moment desire to head into the direction of estimated earliest arrival instead of the direction of the shortest path. This method has been introduced in (Kretz, et al., 2011), (Kretz, et al., 2011), and (Kretz, 2014) as "dynamic potential".

|  | **Before assignment** | **Equilibrium** | **One-shot assignment** |
|---|---|---|---|
| **Average** | 187,1 ± 3,2 s | 163,5 ± 2,4 | 152,3 ± 2,3 |
| **Last** | 335,0 ± 5,8 s | 280,1 ± 5,0 | 263,4 ± 5,8 |

**Table 1: Average time for a pedestrian to reach the destination and time for last pedestrian and standard deviations (in seconds) from 100 simulation runs each.**

Figure 11 shows the density distribution as it occurs within the simulation with one-shot assignment.

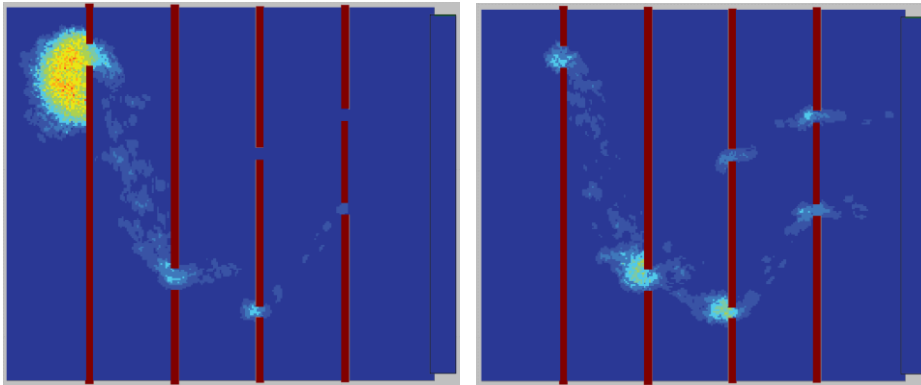

**Figure 11: Density distribution after 50 and after 125 seconds of simulation in a one-shot assignment approach (dynamic potential). It can be seen that – without explicit routes – pedestrians can also in this way be made to utilize both doors where there are two available and it can be seen – particularly at the first door at t=50 s and the second door at t=125 s – that pedestrians distribute better around the door and thus use the available width more efficiently.**

**4. Summary, Conclusions, Discussion, and Outlook**

In this contribution we have summarized the difficulties in simulations of pedestrians to guide pedestrians on – under global perspective – detours without introducing artifacts into their trajectories. We have then sketched the basic elements for a solution which makes use of intermediate destination areas of certain shape at certain location.

The proposed scheme then was applied with an example scenario. It could be demonstrated that even with a very simple assignment method (Equation 1) it is possible to find a travel-time-based equilibrium for the route choice ratios and that the resulting average travel times as well as the evacuation times were significantly smaller for the equilibrium case.

These results in a further step were compared to a simulation with a one-shot assignment approach. In Figure 11 and also in Table 1 it can be seen that the one-shot approach yields the largest overall efficiency in pedestrian behavior. It may well be that this is even more realistic than the approach with explicit intermediate destination (although this remains to be shown empirically). The drawback of the one-shot assignment is, however, as already stated that the route choice ratios are not an explicit result of the method. In a multi-origin and multi-destination scenario they might be difficult to be restored from the data and they might be exactly what a planner desires to have to determine for example the emergency routing in a large and complex infrastructure. It is too early to say how the two methods relate. Particularly as so far they have not been compared in more complex scenarios.